\documentclass[twocolumn]{article}
 
\usepackage{graphicx}

\newdimen\headerboxheight
\newdimen\betweenumberspace          % dimension for space between
\newdimen\aftertext                  % dimension for space after
\newdimen\headlineindent             % dimension for space between
\if@mathematic
   \def\vec#1{\ensuremath{\mathbf{#1}}}
\else
   \def\vec#1{\ensuremath{\mathchoice{\mbox{\boldmath$\displaystyle#1$}}
                              {\mbox{\boldmath$\textstyle#1$}}
                              {\mbox{\boldmath$\scriptstyle#1$}}
                              {\mbox{\boldmath$\scriptscriptstyle#1$}}}}
\fi
\newcommand{\rmi}{\textrm{i}}
\newcommand{\rmd}{\textrm{d}}
\newcommand{\be}{\begin{equation}}
\newcommand{\ee}{\end{equation}}
\newcommand{\beq}{\begin{equation}}
\newcommand{\eeq}{\end{equation}}
\newcommand{\slh}{\!\!\!/} %Dirac slash
\newcommand{\slhh}{\!\!\!\!/} %Dirac slash (for large letters)
\newcommand{\slhhh}{\!\!\!\!\!/} %Dirac slash (even larger)

\begin{document}
 
\twocolumn[

\vspace{3em}

\begin{flushleft}
\small
10 May 2000\\
OSLO-TP 3-00\\
ZAGREB-ZTF-00/02
\end{flushleft}

\vspace*{2em}

\textbf{\textsf{\LARGE 
Bound-state effects in \boldmath $\mu^{+}e^{-} \to \gamma \gamma$
\unboldmath and \boldmath
$\bar{B}^{0}_{s} \to \gamma \gamma$ \unboldmath decays
}}

\vspace{1em}

J. O. Eeg$^{1a}$ , K. Kumeri\v{c}ki$^{2b}$ 
and I. Picek$^{2c}$

\vspace{1em}
{\small
$^1$ Department of Physics, University of Oslo, N-0316 Oslo, Norway\\
$^2$ Department of Physics, Faculty of Science, University of Zagreb, 
 P.O.B. 331, HR-10002 Zagreb, Croatia
}
\vspace{1em}

%\subsection*{Abstract}

\begin{center}
\begin{minipage}{15cm}
{\small \textbf{Abstract.}
We demonstrate that in the double-radiative decays of heavy-light
QED and QCD atoms, $\mu^{+} e^{-} \rightarrow \gamma\gamma$ and
$\bar{B}^{0}_{s} \rightarrow\gamma\gamma$, there is a contribution coming from
operators that vanish on the free-quark mass shell.  This off-shell
effect is suppressed with respect to the effect of the well known
flavour-changing magnetic-moment operator by the bound-state binding
factor. Accordingly, the negligible off-shellness of the weakly bound
QED atoms becomes important for strongly bound QCD atoms. We present
this effect in two different model-approaches to QCD, one of them
enabling us to keep close contact to the related effect in QED.
}
\end{minipage}
\end{center}

\vspace*{1em}

%\vskip0.5cm\hrule\vskip3ptplus12pt\null
]
\renewcommand{\thefootnote}{\alph{footnote}}
\stepcounter{footnote}
\footnotetext{j.o.eeg@fys.uio.no}
\stepcounter{footnote}
\footnotetext{kkumer@phy.hr}
\stepcounter{footnote}
\footnotetext{picek@phy.hr}
\renewcommand{\thefootnote}{\arabic{footnote}}
\setcounter{footnote}{0}
% END TITLE
\section{Introduction}

In this paper we focus on the particular off-shell (or the binding) effects 
in the heavy-light fermion systems, common to QED and QCD. 
Such a comparative study
throws  light on the off-shell nonperturbative effects of
valence quarks, studied first by two of us for the double  radiative
decays of the $K_L$ \cite{EeP93,EeP94} and $B_{s}$ meson \cite{EeP94b}.
Subsequently, this study has been continued within the specific bound state
models, both for $K_L \rightarrow 2\gamma$ \cite{KeKKP95} and for 
$\bar{B}^{0}_{s}  \rightarrow 2\gamma$ \cite{VaE98}.
In these papers it was explicitly demonstrated that operators that
vanish by using the perturbative equations of motion gave nonzero 
contributions for processes involving bound quarks. One of
the purposes of the present paper is to demonstrate similar effects for the
bound leptons.
 
To be specific, we consider such 
off-shell effects for two-photon annihilation of the $\mu^+ e^-$ atom,
called {\em muonium}. The off-shell effects will be given
in terms of the binding factor characterizing a given
bound state.
The role of this binding factor becomes more transparent
in the case of the radiative  decay of such a QED atom, where one deals
with the simple Coulomb binding. This
enables us to clearly demonstrate the  off-shell effect in the QED 
case.

A careful study of these effects 
is motivated by the suitability of 
both lepton-changing transition $\mu \rightarrow  e \gamma \gamma$, 
and $\bar{B}^{0}_{s}\rightarrow\gamma\gamma$ decay, to test the
 standard model (SM) and to infer on the 
physics beyond the standard model (BSM).

By selecting the heavy-light muonium system $\mu^+ e^-$ 
(where $m_{\mu}\equiv M \gg m_e \equiv m$), the bound-state calculation 
corresponds to that of the relativistic hydrogen. Thereby we distinguish
between the Coulomb field responsible for the binding, and the radiation field
\cite{Sa67}
participating in the flavour-changing transition at the pertinent 
high-energy scale.
In this way the radiative disintegration of an atom becomes tractable 
 by implementing the two-step treatment \cite{ItZ80}:
``neglecting at first annihilation to compute the binding and then neglecting
binding to compute annihilation''.
For the muonium atom at hand, the binding problem     
is analogous to a solved problem of the H-atom. 
In this way we  avoid the 
relativistic bound state problem, which is a difficult
subject, and we have no intention to contribute to it here.

This two-step method is known to work well for 
 disintegration (annihilation) of the simplest
QED atom, positronium.
Generalization of this procedure to muonium means  that the two-photon
decay width of muonium is obtained by using
\begin{equation}
\Gamma=\frac{|\psi(0)|^2\: |{\cal M}(\mu^{+}e^{-}\to\gamma\gamma)|^2}
            {64 \pi M m} \;,
\label{width}
\end{equation}
where $|\psi(0)|^2$ is the square of the 
bound-state wave function at the origin. 
After this factorization has been performed the rest of the problem
reduces to the evaluation of the scattering-annihilation invariant
amplitude ${\cal M}$.
In the case of positronium this expression will involve equal
masses ($M$=$m$), and 
  the invariant
amplitude which for the positronium annihilation at rest has a textbook
form~\cite{Na90}
\begin{equation}
{\cal M}=\frac{\rmi e^2}{2m^2}\bar{v}_{s}(p_2)\Big\{
\epsilon\slh_{2}^{*}\epsilon\slh_{1}^{*}k\slh_{1}+
\epsilon\slh_{1}^{*}\epsilon\slh_{2}^{*}k\slh_{2}
\Big\}
   u_r(p_1) \;.
\label{positronium}
\end{equation}
Only the antisymmetric piece in the decomposition of the product of
three gamma matrices above
\begin{equation}
\Big\{ \;\; \Big\} \to \rmi
 \epsilon ^{\mu \nu \alpha \beta} \gamma_5 \gamma_{\beta}
(k_1 - k_2)_{\alpha}
(\epsilon ^*_1)_{\mu}(\epsilon ^*_2)_{\nu} \;,
\label{parapositronium}
\end{equation}
contributes to the spin singlet parapositronium two-pho\-ton 
annihilation. This selects
($\vec{\epsilon}^{\ast}_{1}\times\vec{\epsilon}^{\ast}_{2}$),
a CP-odd configuration 
of the final two-photon state.

If parapositronium decay can serve as an initial benchmark in considering
QED atom annihilation, then its QCD counterpart would be  
$\pi^{0}\to\gamma\gamma$. However, the latter process shows some
subtlety, known as the triangle anomaly.
Interestingly enough, this quark atom double radiative decay 
can also be viewed as an off-shell effect, 
as explained in some detail in \cite{EeP94c}. 
It is the off-shellness in two-photon annihilation of atoms which we
further explore in what follows.

The paper is organized as follows: In section 2 we we consider the
quantum field treatment of the annihilation process
$\mu^{+}e^{-} \to \gamma \gamma$  in 
 arbitrary external field(s).
In section 3 
we relate the binding forces to the external fields of section 2.  In
section 4 we consider the analogous heavy-light QCD system, and in
section 5 we give our conclusions.

\section{Flavour-changing operators for $\mu^{+}e^{-} \to \gamma \gamma$}

\begin{figure}
\centerline{\includegraphics[scale=0.8]{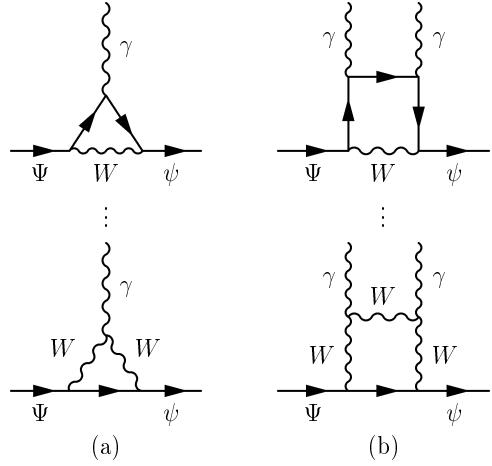}}
\caption{The examples of the one-particle-irreducible diagrams leading to the
 double-radiative flavour-changing transitions. Only the second-row diagrams
exist for the leptonic case.
\label{1PI}}
\end{figure}

We treat the lepton flavour-changing process at hand analogously to
the quark flavour change, accounted for by the electroweak theory. Thus,
the double-radiative 
transition is triggered by two classes of one-particle-irreducible
diagrams (Fig. \ref{1PI}a and Fig. \ref{1PI}b), 
related by the Ward identities. 
After integrating out the heavy  particles in the loops, 
these one-loop electroweak  transitions can be combined  into
an effective Lagrangian \cite{EeP93},
\beq
{\cal{L}}(e \rightarrow \mu)_{\gamma} \, = \,  B \, 
\epsilon^{\mu \nu \lambda \rho}
F_{\mu \nu} \, ( \bar{\Psi} \; \rmi \stackrel{\leftrightarrow}{D_{\lambda}}
 \gamma_{\rho} L\, \psi )  \, + \mbox{h.c.}  \; .
\label{eq:lagg}
\eeq
 where 
the muon and the electron
 are described by quantum fields $\Psi = \psi_{\mu}$ and $\psi= \psi_e$.
Correspondingly, for $\bar{B}^{0}_{s} \rightarrow 2 \gamma$, the involved fields
are $\psi_s=s$ and $\psi_b=b$.

In our case, we do not need to specify the physics behind the 
lepton-flavour-violating  transition in (\ref{eq:lagg}). 
For instance, the strength $B$ might contain  some leptonic parameters, 
analogous to the Cabibbo-Kobayashi-Mas\-ka\-wa parameters $\lambda_{\rm CKM}$
in the quark sector.

Keeping in mind that the fermions in the bound states are not
on-shell, we are not simplifying the result of the electroweak
loop calculation by using the perturbative eq\-uation of motion.
Thus the effective Lagrangian (\ref{eq:lagg})
obtained within perturbation theory 
splits into the on-shell magnetic transition operator ${\cal{L}}_{\sigma}$
\beq
{\cal{L}}_{\sigma} (1 \gamma) = B_{\sigma} \bar{\Psi} 
 \, (M  \sigma \cdot F L + 
 m\,  \sigma\cdot F R) \,  \psi \, + \mbox{h.c.} \; , 
\label{eq:lagmag}
\eeq
and an off-shell piece ${\cal{L}}_F$ \cite{EeP93} 
\beq
{\cal{L}}_{F} = B_F\, \bar{\Psi} [(\rmi \stackrel{\leftarrow}{D}\slhhh - M) \,
 \sigma \cdot F L + \sigma \cdot F R (\rmi
D\slhh -m)] \psi \, + \mbox{h.c.} \; , 
\label{eq:laggg}
\eeq
where $\sigma\cdot F$ denotes $\sigma_{\mu \nu} F^{\mu \nu}$, and
$L=(1-\gamma_5)/2$ and $R=(1+\gamma_5)/2$ denote left-hand and right-hand
projectors.
To lowest order in QED (or QCD) $B_F = B_\sigma = B$, but in general they are
different due to different anomalous dimensions of the operators in
(\ref{eq:lagmag}) and (\ref{eq:laggg}). (The off-shell part ${\cal{L}}_{F}$ 
has zero anomalous dimension).

\begin{figure}
\centerline{\includegraphics[scale=0.8]{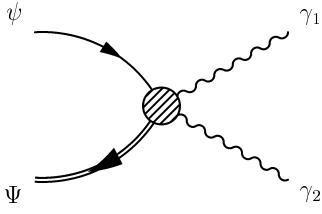}}
\caption{The two-photon contact (seagull) diagram that can be rotated
  away by a field redefinition.\label{seagull}}
\end{figure}

By decomposing the covariant derivative, 
$\rmi D\slhh =  \rmi \partial\slh - e A\slh$, in the off-shell operator
(\ref{eq:laggg}), we separate the one-photon piece
\beq
{\cal{L}}_F(1 \gamma)
 = B_F \, \bar{\Psi} 
[(\rmi \stackrel{\leftarrow}{\partial\slh} - M) \,
 \sigma \cdot F L + \sigma \cdot F R (\rmi
\partial\slh -m)] \psi \, + \mbox{h.c.}\; ,
\label{eq:lag1g}
\eeq
from the two-photon piece
\beq
{\cal{L}}_F(2  \gamma)
= B_F\, \bar{\Psi} [- e A\slh \sigma \cdot F L +
\sigma \cdot F R (- e A\slh)]\psi \, + \mbox{h.c.}\; .
\label{eq:lag2g}
\eeq
The amplitude for the two-photon diagram (Fig. \ref{seagull}) is given by
\beq
A_a = \rmi \int \rmd^4x \, {\cal{L}}_F(2  \gamma)
 = A^L_a + A^R_a \;,
\label{eq:lag2gD}
\eeq
in an obvious notation. 
The single-photon off-shell Lagrangian ${\cal{L}}_F(1 \gamma)$
leads to the amplitude with the heavy particle in the propagator
\begin{eqnarray}
A_b &=&  \rmi \,  B_F \, \int \!\! \int \rmd^4x \, \rmd^4y \, \bar{\Psi}(y) 
\Big[-\rmi e  A\slh_{2}(y)\Big]\, \rmi S_F^{(\mu)}(y-x) \nonumber \\
&&\hspace*{-2em}\times\bigg[(\rmi \stackrel{\leftarrow}{\partial\slh_x} - M) \,
 \sigma \cdot  F_1(x) L   + \sigma \cdot  F_1(x) R 
  (\rmi \partial\slh_x -m)\bigg] \psi(x) \; , \nonumber \\
\label{eq:lag1gDb}
\end{eqnarray}
and a similar amplitude with the light particle in the propagator
\begin{eqnarray}
A_c & = &  \rmi \,  B_F \, \int \!\! \int \rmd^4x \,\rmd^4y \, \bar{\Psi}(x) 
 \bigg[(\rmi \stackrel{\leftarrow}{\partial\slh_x} - M) \,
 \sigma \cdot  F_1(x) L +  \nonumber \\
&& \hspace*{-2em} \sigma \cdot  F_1(x) R (\rmi
\partial\slh_x -m)\bigg]  
\times \, \rmi S_F^{(e)}(x-y) \Big[-\rmi e A\slh_2(y)\Big] 
\psi(y) \; . \nonumber \\
\label{eq:lag1gDc}
\end{eqnarray}
The subscripts 1 and 2 distinguish between the two photons. It is
understood that a term with the $1 \leftrightarrow 2$ subscript
interchange should be added in order to make our result symmetric in the
two photons.

Within the quantum field formalism, the sum of the equations
(\ref{eq:lag2gD}), (\ref{eq:lag1gDb}) and (\ref{eq:lag1gDc})
describes the process
$\mu^{+}e^{-} \to \gamma \gamma$, or 
$\mu  \to  e \gamma \gamma$.

Let us now be very general, and assume that both particles
($e$ and $\mu$) feel some kind of external field(s)  represented by
$V_{(e)}$ and $V_{(\mu)}$, and obey 
 one-body Dirac equations
\begin{equation}                                                               
  \big[ \rmi \partial\slh - V_{(i)}(x) - m_{(i)} \big] \psi_{(i)} = 0 \;,
\label{CoulDir}                                                                
\end{equation}
for $i= e$ or  $\mu$ (in general $V_{(i)}= \gamma_{\alpha} \, V^{\alpha}_{(i)}$)
, and accordingly
the particle pro\-pa\-ga\-tors
$S_F^{(i)}$ satisfy:
\begin{equation}                                                               
  \big[ \rmi \partial\slh - V_{(i)}(x) - m_i \big] S^{(i)}_{F}(x-y)=                  
 \delta^{(4)}(x-y) \;.                                                         
\label{lightprop}                                                              
\end{equation}
Our photon fields enter via
perturbative  QED, switched on by the replacement
$\partial_\mu \rightarrow D_\mu = \partial_\mu + i e A_\mu$ 
in (\ref{CoulDir}).
 It should be emphasized that $A_{\mu}(x)$ represents the radiation field and
does not include binding forces, which will in the next section 
be related to the external fields $V_{(i)}$.

Now, using relations (\ref{CoulDir}) and (\ref{lightprop}) we obtain
\beq
A_b  = -A_a^L + \Delta A_b \; , \qquad A_c  = -A_a^R + \Delta A_c  \;,
\eeq
resulting in a partial cancellation when the amplitudes are summed
\beq
A_a + A_b  + A_c =  \Delta A_b + \Delta A_c \; .
\eeq
This shows that the local off-diagonal fermion seagull 
transition of Fig. \ref{seagull} cancels, even if the external
fermions are off-shell.
The left-over  quantities $\Delta A_b$ and  $\Delta A_c$ involve 
the integrals over the Coulomb potential and represent the net
off-shell effect.

\begin{figure}
\centerline{\includegraphics[scale=0.7]{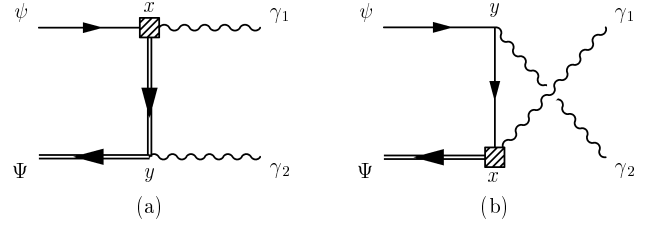}}
\caption{The shaded boxes indicate the combination of the
unrotated off-shell transition ($\sim B_F$) and the on-shell magnetic
moment transition ($\sim B_{\sigma}$), giving the effective vertex
in Eq. (\ref{shaded}).\label{MbMc}}
\end{figure}

There are also amplitudes $A_d$ and $A_e$ which are counterparts 
of $A_b$ and $A_c$
when ${\cal{L}}_F(1 \gamma)$ is replaced by ${\cal{L}}_{\sigma}$.
The total contribution from our flavour-changing Lagrangian 
(${\cal L}_F$ and ${\cal L}_\sigma$ parts) is then given by
\begin{eqnarray}
A_d + \Delta A_b & = &
\, \rmi    \int \!\! \int \rmd^4x\, \rmd^4y\,  \bar{\Psi}(y) 
\Big[-\rmi e A\slh_2(y)\Big] \hspace*{5em} \nonumber \\
& &\hspace*{5em}{} \times  \rmi S_F^{(\mu)}(y-x) \, Q(x) \, \psi(x) \; , 
\label{eq:lag1gDd}
\end{eqnarray}
represented by Fig. \ref{MbMc}a, and a similar one
\begin{eqnarray}
A_e + \Delta A_c  & = &
\,  \rmi   \int \!\! \int \rmd^4x\, \rmd^4y \, \bar{\Psi}(x) \, Q(x) \,    
 \rmi S_F^{(e)}(x-y)  \hspace*{3em} \nonumber \\
& & \hspace*{5em}{} \times  \Big[-\rmi e A\slh_2(y)\Big]\psi(y) \; ,
\label{eq:lag1gDe}
\end{eqnarray}
corresponding to Fig. \ref{MbMc}b. The operator $Q(x)$ in these expressions 
reads
\begin{eqnarray} 
\lefteqn{Q(x)  =   \big[ B_{\sigma}M + B_F V_{(\mu)}(x) \big] \,
  \sigma \cdot  F_1(x) L } \hspace*{5em}  \nonumber \\
& & {}+ \sigma \cdot  F_1(x) R \, \big[ B_{\sigma}m + B_F V_{(e)}(x) \big]\,.
\label{eq:Qexpr}
\end{eqnarray}
The result given by Eqs. (\ref{eq:lag1gDd})--(\ref{eq:Qexpr}) 
can also be understood in terms of the following
field redefinition.
Eq. (\ref{CoulDir}) can be obtained from the Lagrangian
\begin{equation}
{\cal{L}}_{D}(\Psi,\psi) = 
\bar{\Psi} \big[ \rmi D\slhh - V_{(\mu)} - M \big] \Psi +
\bar{\psi} \big[ \rmi D\slhh - V_{(e)} - m \big] \psi \; . 
\label{AppDir}
\end{equation}
Now, by defining new fields
\beq
\Psi ' = \Psi + B_F\,  \sigma \cdot  F \,  L  \psi \; , \qquad 
\psi ' = \psi + B_F ^* \,  \sigma  \cdot F  L \Psi \; ,
\label{repsi}
\eeq
we obtain
\beq
{\cal{L}}_{D}(\psi,\Psi) +  {\cal{L}}_{F}  \, = \,
{\cal{L}}_{D}(\psi ' , \Psi ') \, + \Delta {\cal{L}}_{B} \;,
\label{Trlag}
\eeq
which shows that ${\cal{L}}_{F}$ can be transformed away from 
the perturbative terms,
but a relic of it,
\beq
\Delta  {\cal{L}}_{B} \, = \, 
  B_F\, \bar{\Psi} \big[ V_{(\mu)} \sigma \cdot  F  L  \, + \,
  \sigma \cdot F R \,  V_{(e)} \big] \psi  \, + \textrm{h.c.} \; , 
\label{newlag}
\eeq
remains in the bound state dynamics . Thus, the off-shell effects
are non-zero for bound external fermions. Combining
$\Delta {\cal{L}}_{B}$ and ${\cal{L}}_{\sigma}$, we obtain
\beq
\Delta {\cal{L}}_{B} \, + \,{\cal{L}}_{\sigma} \; = \; 
\bar{\Psi} \, Q \, \psi \, + \textrm{h.c.} \; ,
\label{TotB}
\eeq
where $Q$ is given by (\ref{eq:Qexpr}). This shows the equivalence
of this field redefinition procedure and the result
given by Eqs. (\ref{eq:lag1gDd})--(\ref{eq:Qexpr}).

This is how far we can push the problem within quantum field theory.
Up to now we have made no approximations except for standard 
perturbation theory. In the next section we will  adapt the results
of this section to the relevant bound state effects.

\section{Off-shellness in the muonium annihilation amplitude}

As announced in the Introduction, we choose the simplest
heavy-light QED atom,
muonium. 
Naively, the product $\bar{\Psi} \, \psi$ corresponds to the bound state
of $\mu^+$ and $e^-$  (which might be true only in the asymptotically free
case). However, relativistic bound state physics is a difficult
 subject, out of our scope.
We will stick to the two-step procedure \cite{ItZ80}
 as explained in the Introduction.
We perform the calculations in the muonium rest frame (CM frame of
 $\mu^+$ and $e^-$) where we put the external field(s) equal
to a mutual Coulomb field, $V_{(i)} \rightarrow \gamma_0 \, V_C$
(where $V_C = -e^2/4 \pi r$). In calculating the
$\mu^+ e^- \rightarrow \gamma \gamma$ amplitude in momentum space,
we take for $V_C$ the 
 average over  solutions in the
Coulomb potential,
 which is
$\langle V_C\rangle = - (m\alpha^2/2)$.
 In this way the 
muonium-decay invariant amplitude acquires the form which is
a straightforward generalization of the positronium-decay
invariant amplitude (\ref{positronium}) in momentum space.

The amplitudes $A_d + \Delta A_b$ from Eq. (\ref{eq:lag1gDd}), together
with $A_e + \Delta A_c$ from (\ref{eq:lag1gDe}), transformed to
the momentum space take the form
\begin{eqnarray}
{\cal M}&=&\frac{2 e B_{\sigma}}{m}\bar{v}_{\mu}(p_2)\Big\{
\frac{m}{M}k\slh_{2}\epsilon\slh_{2}^{*}{\rm P}-
{\rm P}\epsilon\slh_{2}^{*}k\slh_{2} + (1\leftrightarrow 2)
\Big\}
   u_{e}(p_1), \nonumber \\
\label{5slashes}
\end{eqnarray}
where $v_{\mu}$ and $u_{e}$ are muon and electron spinors, and 
$\epsilon_{1,2}^{*}$ are photon polarization vectors.
The factor, incorporating the binding in the form of a four-vector
$U^{\alpha}=(\rho,\vec{0})$,
\begin{eqnarray}
{\rm P}\equiv(1-xU\slhh)k\slh_{1}\epsilon\slh_{1}^{*}L +
         xk\slh_{1}\epsilon\slh_{1}^{*}R(1-U\slhh) \;,
\end{eqnarray}
accounts for the aforementioned factorization of a binding and a decay,
and is represented by the shaded box of Fig. \ref{MbMc}:
\begin{equation}
 \Bigg[ M(1-x \rho \gamma^{0})\sigma \cdot F_{1}L + 
 m \,\sigma \cdot F_{1}R(1-\rho \gamma^{0}) \Bigg] \;.
\label{shaded}
\end{equation}
Here we introduced abbreviations for two small constant parameters,
\begin{equation}
x \equiv \frac{m}{M} \;, \qquad 
\rho \equiv - \, \frac{B_{F} \, \langle V_C \rangle}{m B_{\sigma}} \;,
\end{equation}
in terms of which the sought off-shell effect will be expressed.
Note that in the effective interaction (\ref{shaded}), the left-handed
part corresponding to $V_{(\mu)}$ has gotten an extra suppression
factor $x=m/M$ in front of the binding factor $\rho$, in agreement with
the expectation that the heavy particle ($\mu^+$) is approximately free,
and the light particle ($e^-$) is approximately the reduced particle,
in analogy with the H-atom.

The annihilation amplitude (\ref{5slashes})
can now be evaluated explicitly. A tedious calculation, performed
 in the muonium rest frame with photons emitted
along the $z$-axis, gives
\begin{eqnarray}
{\cal M} &=& - 2 e B_{\sigma}  M^2 \sqrt{\frac{2M}{m}} \Big[
 (1+x \rho)\vec{\epsilon}_{2}^{*}\cdot\vec{\epsilon}_{1}^{*} \nonumber \\
&& \hspace*{-1em}{}+\rmi(1+2x+x \rho)(\vec{\epsilon}_{2}^{*}
  \times\vec{\epsilon}_{1}^{*})
\cdot\hat{\vec{k}}_{1} + 
  {\cal O}({\rho}^2, x^2) \Big] \;,
\label{Mamp}
\end{eqnarray}
where we have kept only the leading terms in $\rho$ and $x$.
In comparison to the expressions (\ref{positronium}) and
(\ref{parapositronium}) for parapositronium,  we notice that in addition to
$\vec{\epsilon}_{2}^{*}\times\vec{\epsilon}_{1}^{*}$ there appears also
$\vec{\epsilon}_{2}^{*}\cdot\vec{\epsilon}_{1}^{*}$, a CP-even two-photon
configuration.

The explicit expression for $\rho$ depends on some assumptions.
As explained previously, we use $\langle V_C \rangle = -m\alpha^2/2$ 
which gives
$\rho = \alpha^2/2$ for $B_\sigma = B_F = B$, which is a good 
approximation in the leptonic case.

\noindent
Eq.(\ref{width}) finally gives
\begin{equation}
\Gamma = \frac{2 \alpha  M^4}{m^2}|\psi(0)|^2 |B_{\sigma}|^{2}
  \left(1+ 2 x \rho \right) \;.
\label{totalG}
\end{equation}
Thus, for muonium, the sought off-shell contribution is only a tiny 
correction,
$2 x \rho = \alpha^2 m/M\simeq 2.6 \cdot 10^{-7}$, to the magnetic
moment dominated rate\footnote{Note that it is not necessary to know 
the precise value of $|\psi(0)|^2 \sim (m\alpha)^3/\pi$, in order to 
know the relative off-shell contribution.}.
However, the
corresponding off-shellness in a strongly bound QCD system
should be significantly larger.
We also take into account the
$B_{F}/B_{\sigma}$ correction in (\ref{totalG}),
when considering the $\bar{B}^{0}_{s}\to\gamma\gamma$ decay below.

Before ending this section, we should also mention that the Lagrangian
given by (\ref{eq:Qexpr}) and (\ref{TotB}) can be used  
to calculate the amplitude for muonic hydrogen
decaying to a photon and ordinary hydrogen, that is, the process 
$\mu^- \rightarrow e^- + \gamma$ for both leptons bound to
a proton. This is a leptonic version of the celebrated $B$-meson decay
$B_d \rightarrow K^* \gamma $.
 
As a toy model, one might consider a process
 ``$\mu$'' $\rightarrow$ ``$e$'' $\gamma$ in an external Coulomb field,
with ``$\mu$'' and ``$e$'' rather close in mass such that the
non-relativistic descriptions of the ``leptons'' might be used.
The effective ``$\mu$'' $\rightarrow$ ``$e$'' $\gamma$ interaction is
given in (\ref{TotB}). If we assume that $(M-m)$ is of order $\alpha \, m$,
we obtain  off-shell effects of order $\alpha^2$
due to ${\cal{L}}_F$, relative to the standard magnetic moment
term ${\cal{L}}_\sigma$.
Bigger mass 
differences gives bigger effects, until the 
non-relativistic approximation breaks down.

\section{Off-shellness in $\bar{B}^{0}_{s} \to \gamma \gamma$}

By the replacements $\mu \rightarrow s$ and $e \rightarrow b$, the expressions
(\ref{eq:lagg}) to (\ref{eq:lag2g}) apply to  $b\rightarrow s \gamma \gamma$ 
induced $\bar{B}^{0}_{s}\rightarrow 2 \gamma$ decay amplitude. Then one has
to scale the operators ${\cal{L}}_{F,\sigma}$ defined at the $M_{W}$ scale, 
down to the $B$-meson scale. 
The coefficients 
  $B_F$ of ${\cal{L}}_{F}$, and $B_{\sigma}$ of  ${\cal{L}}_{\sigma}$,
in Eqs. (\ref{eq:laggg}) and (\ref{eq:lagmag}),
 both being
equal to $B$ at the $W$ scale, may evolve differently down to the
 $\mu = m_b$ scale.
This difference between $B_{F}$ and $B_{\sigma}$ is due to different anomalous 
dimensions of the respective operators.
Within the SM one can write
\beq
B_{\sigma,F} = \frac{4 G_F}{\sqrt{2}} \, \lambda_{\rm CKM} \,
  \frac{e}{16 \pi^2} \, C_7^{\sigma,F} \; .
\label{defhat}
\eeq
The coefficient $C_7^\sigma$ has been studied by various authors
\cite{GrSW90,BuMMP94,GrOSN88,CeCRV90,CeCRV94}. 
The coefficient $C_7^F$ was considered in \cite{EeP94b}, where at
 the $b$-quark scale we obtained
\beq
\frac{C_7^{F}}{C_7^{\sigma}} \simeq 4/3   
\qquad  (\mu =m_{b}) \; .
\label{Bratio}
\eeq
Although the off-shell effect for $B \rightarrow 2 \gamma$ 
is expected to be suppressed by the ratio (binding
energy)/$m_{b}$, it could still be numerically interesting.

\subsection{Coulomb-type QCD model}

The conventional procedure when evaluating the pseudoscalar meson
decay amplitudes is to express them in terms of the meson decay constants, 
by using the PCAC relations
\begin{eqnarray}
\langle 0 | \bar{s} \gamma_{\mu} \gamma_{5} b | \bar{B}^{0}_{s}(P) \rangle &=& 
- {\rmi f_{B} P_{\mu}} \;, \label{PCAC} \\
\langle 0 | \bar{s}  \gamma_{5} b | \bar{B}^{0}_{s}(P) \rangle  & = &
{\rmi f_{B} M_{B}} \;.
\label{Axcurrent}
\end{eqnarray}
These relations will be useful after reducing our 
general expression (\ref{5slashes})
containing the terms with products of up to five gamma matrices. After
some calculation we arrive at the expression for the $B_s$ meson decay
at rest, which is analogous to, and in fact confirms our previous
relation (\ref{Mamp}) obtained in different way,
\begin{eqnarray}
{\cal M}^{B} &=& - \rmi \frac{e}{3} B_{\sigma} f_{B} M^2 
\frac{(1+x)^2}{x} \Big[
 (1+x \tau)\,\vec{\epsilon}_{2}^{*}\cdot\vec{\epsilon}_{1}^{*} 
  + \nonumber \\
 & + & \; \rmi(1+2x+x \tau)(\vec{\epsilon}_{2}^{*}
  \times\vec{\epsilon}_{1}^{*})
\cdot\hat{\vec{k}}_{1} + {\cal O}(\tau^2, x^2) \Big].
\label{Mampb}
\end{eqnarray}
Here, the parameter $\tau$ represents the off-shell effect in the
QCD problem at hand,
and will be more model dependent than its QED counterpart $\rho$.
With the amplitude (\ref{Mampb})
we arrive at the total decay width
\begin{equation}
\Gamma = \frac{\alpha  M^5}{18 m^2} f_{B}^{2}
 |B_{\sigma}|^2 \left(1+ 2 x \tau \right) \;,
\end{equation}
where by switching off  $\tau$ we reproduce the result of Ref.
\cite{DeDL96}.

In order to estimate the value of the off-shell contribution $\tau$, in
this subsection we assume a QED-like QCD model with the Coulombic wave
function \cite{Go97,TrM88}
$\psi(r)\sim \exp(-m r \alpha_{\rm eff})$.
Thus we rely again on an exact solution cor\-res\-pon\-ding to effective 
potential
$V(r) = - 4\alpha_{\rm eff}/(3 r)$, with effective coupling 
$\alpha_{\rm eff}(r)\! = \! -(4\pi b_{0} \ln (r\Lambda_{\rm pot}))^{-1}$. 
%\\
Here $b_0=(1/8\pi^2)(11-(2/3)N_f)$.  The mass scale $\Lambda_{\rm pot}$
appropriate to the heavy-light quark $\bar{Q}q$ potential
 is related to the more familiar
QCD scale parameter, e.g. $\Lambda_{\rm pot}=2.23\,
  \Lambda_{\rm \overline{MS}}$
(for $N_f$=3). Within this model, we obtain
\begin{equation}
\tau =\frac{2}{3}\alpha_{\rm eff}^2 \frac{C_{7}^{F}}{C_{7}^{\sigma}}
\;.
\end{equation}
By matching the meson decay constant
$f_{B}$ and the wave function at the origin 
\begin{equation}
N_{c}\frac{|\psi_{B}(0)|^2}{M}=\left(\frac{f_B}{2}\right)^2 \; ; \quad
|\psi_{B}(0)|^2=\frac{(m \alpha_{\rm eff})^3}{\pi} \;,
\end{equation}
we obtain the value for the strong interaction fine structure 
strength $\alpha_{\rm eff}$ $\sim$1.
Then, including (\ref{Bratio}) for the QCD case, the correction
factor
\begin{equation}
 x \tau \simeq\, 0.1\,  \;,
\label{QCDoff}
\end{equation}
is much larger than $x\rho$ in the corresponding QED case.
Correspondingly, one expects even more significant off-shell effects
in light quark systems, in compliance with our previous
results \cite{EeP93,EeP94,KeKKP95}.

\subsection{A constituent  quark  calculation}

Now we adopt a variant of the approach in Refs. \cite{EeP94b,VaE98} as
an alternative to the Coulomb-type QCD model described
above. One might use the PCAC relations 
(\ref{PCAC})--(\ref{Axcurrent}) together with a kinematical
assumption for the $\bar{s}$-quark momentum, similar to those in Refs.
\cite{DeDL96,HeK92}. Assuming the bound $\bar{s}$ and
$b$ quarks in $\bar{B}^{0}_s$ to be on their respective \emph{effective} mass-shells
(effective mass being current mass plus a constituent mass $m_0$ of order
200-300 MeV), the structure of the amplitude comes out essentially as in
(\ref{Mampb}) with a relative off-shell contribution
\begin{equation}
x \, \tilde{\tau} \, = \, \frac{2 m_0}{m_b} \, \sim 0.1 \; ,
\label{Bsoff}
\end{equation}
of the same order as in (\ref{QCDoff}). However, unlike (\ref{Mampb}), the
off-shell effect is now only in the CP-odd term 
($\vec{\epsilon}^{\ast}_{1}\times\vec{\epsilon}^{\ast}_{2}$), the
square bracket in (\ref{Mampb}) being replaced by
\begin{equation}
\Big[\vec{\epsilon}_{2}^{*}\cdot\vec{\epsilon}_{1}^{*} 
  + 
 \rmi(1+2x+x \tilde{\tau})(\vec{\epsilon}_{2}^{*}
  \times\vec{\epsilon}_{1}^{*}) \Big] \;.
\end{equation}
This may
be different in other approaches \cite{EeP94}, showing the model dependence
of the off-shell effect. For instance, potential-QCD models in general, besides
a vector Co\-u\-lomb potential, also contain a scalar potential.

\subsection{A bound state quark  model}

For $\bar{B}^{0}_{s} \rightarrow 2 \gamma$, we have previously 
\cite{EeP94b,VaE98} 
 applied  a bound state model, where the potentials $V_{i}$ in
 (\ref{CoulDir}) are  replaced by a quark-meson interaction 
Lagrangian
\beq
{\cal{L}}_{\Phi}(s,b) \, = 
\, G_B \bar{b} \, \gamma_5 \, s \, \Phi + 
  {\rm h.c.} \; ,
\label{qmes}
\eeq
where $\Phi$ is the B-meson field. Then, the term ${\cal{L}}_{F}$ can be
 transformed away by means of the  field redefinitions:
\beq
 s' = s + B_F\,  \sigma \cdot  F \,  L  \, b \; , \qquad 
 b' = b + B_F ^* \,  \sigma  \cdot F  L \, s \; .
\label{rebs}
\eeq
However, its 
 effect  reappears in a new bound-state
 interaction $\Delta {\cal{L}}_{\Phi}$,
\beq
{\cal{L}}_{\Phi}(s,b)   + {\cal{L}}_{F} \, = \,
{\cal{L}}_{\Phi}(s',b') \, + \Delta {\cal{L}}_{\Phi} \;,
\label{DeltaL}
\eeq
where, after using $R \gamma_5 = R$ and $L \gamma_5 = -L$,
\beq
\Delta  {\cal{L}}_{\Phi} \, = \, 
  B_F \, G_B \,  \big[ \bar{b}' \sigma \cdot  F  L  \, b' \, - \,
  \bar{s}'  \, \sigma \cdot F R \, s'  \big] \Phi   \, + {\rm h.c.} \;.
\label{newInt}
\eeq
Also in these cases \cite{EeP94b,VaE98}, the net off-shell effects are found.
Further calculations of  $B \rightarrow 2\gamma$ within bound state
models of the type in (\ref{qmes}) will be presented elsewhere.

\section{Conclusions}

We have demonstrated the appearance of the
off-shell effects in the flavour-changing 
two-photon decay of muonium  and its 
hadronic $\bar{B}^{0}_{s} \rightarrow \gamma
 \gamma$ counterpart. It is a quite significant 10 percent effect in the 
latter case, whereas in the leptonic case
it is very small (of order 10$^{-7}$), but clearly identifiable.

The present ``atomic'' approach enables us to see in a new light the effect 
studied first for the $K_L\rightarrow  \gamma\gamma$ amplitude in the chiral 
quark model \cite{EeP93,EeP94}, and subsequently in the bound-state model 
\cite{KeKKP95}. The observation that off-shell effects can be clearly isolated 
from the rest in the heavy-light quark atoms \cite{EeP94b} was still plagued 
by the uncertainty in the QCD binding calculation \cite{VaE98}. 
Here, in the Coulomb-type QCD model we are able to subsume the effect into 
an universal binding factor, in the same way as for the two-photon decay of 
muonium in the exactly solvable QED framework.
As a result, we obtain the explicit expressions describing how
the flavour-changing operators that vanish on-shell modify both the
CP-even ($\vec{\epsilon}^{\ast}_{1}\cdot\vec{\epsilon}^{\ast}_{2}$), and  
CP-odd ($\vec{\epsilon}^{\ast}_{1}\times\vec{\epsilon}^{\ast}_{2}$)
configuration of the final photons.
In a constituent quark calculation we get a similar result for
the off-shell effect. As a difference, in this case the off-shellness resides
solely in the CP-odd part of the amplitude.

The main result of the present paper is a clear demonstration of
the parallelism of the strict nonzero off-shell effects in the leptonic
and quark heavy-light systems. Thus, we have established
a solid ground for estimating the off-shell bound-state effects
in the important $B_{d}\to K^{*}\gamma$ decay, which will be presented
elsewhere \cite{EKPprep}.

\vspace*{1em}
\small
\noindent
\textbf{Acknowledgement.}
One of us (J.O.E.) thanks H. Pilkuhn for discussions related to this paper.

\end{document}